\pdfoutput=1
\documentclass[aps,pra,showkeys,showpacs,notitlepage,superscriptaddress]{revtex4-1}
\usepackage{amsmath}
\usepackage{amssymb}
\usepackage{graphicx}
\usepackage{subfigure}
\usepackage{enumerate}
\usepackage{todonotes}
\usepackage{xspace}
\usepackage{tikz}
\usetikzlibrary{shapes, calc}
\usepackage{hyperref}
\usepackage{algorithm}
\usepackage[noend]{algpseudocode}

\newcommand{\Tr}{\mathrm{Tr}}
\newcommand{\tr}{\Tr}

\newcommand{\ket}[1]{\ensuremath{|#1\rangle}}

\newcommand{\1}{{\rm 1\hspace{-0.9mm}l}}

\newcommand{\ii}{\mathrm{i}}
\newcommand{\dd}{\mathrm{d}}

\newcommand{\Cplx}{\mathbb{C}}

\newcommand{\Sx}{X}
\newcommand{\Sz}{Z}
\newcommand{\Sy}{Y}

\newcommand{\ie}{\emph{ie.}\xspace}

\newcommand{\qutip}{QuTIP\xspace}

\newcommand{\NCP}{\ensuremath{h^{\gamma=0}}\xspace}
\newcommand{\DCP}{\ensuremath{h^{\gamma >0}}\xspace}
\newcommand{\DCPinA}{\ensuremath{h^{\gamma \in A}}\xspace}
\newcommand{\nnDCPniA}{\ensuremath{\tilde{h}^{\gamma \not\in A}}\xspace}
\newcommand{\CCP}{\ensuremath{\Delta h}\xspace}
\newcommand{\nnDCP}{\ensuremath{\tilde{h}^{\gamma >0}}\xspace}
\newcommand{\kNN}{\ensuremath{\mathrm{kNN}}\xspace}

\newcommand{\mySigma}{\sigma}

\begin{document}

  \title{Geometrical versus time-series representation of data in quantum control learning}

  \author{M. Ostaszewski}
  \affiliation{Institute of Theoretical and Applied Informatics, Polish Academy of
    Sciences, Ba{\l}tycka 5, 44-100 Gliwice, Poland}
  \author{J.A. Miszczak}
  \affiliation{Institute of Theoretical and Applied Informatics, Polish Academy of
    Sciences, Ba{\l}tycka 5, 44-100 Gliwice, Poland}
  \author{P. Sadowski}
  \affiliation{Institute of Theoretical and Applied Informatics, Polish Academy of
    Sciences, Ba{\l}tycka 5, 44-100 Gliwice, Poland}

\begin{abstract}
    Recently machine learning techniques have become popular for analysing physical systems and solving problems occurring in quantum computing. In this paper we focus on using such techniques for finding the sequence of physical operations implementing the given quantum logical operation. In this context we analyse the flexibility of the data representation and compare the applicability of two machine learning approaches based on different representations of data. We demonstrate that the utilization of the geometrical structure of control pulses is sufficient for achieving high-fidelity of the implemented evolution. We also demonstrate that artificial neural networks, unlike geometrical methods, posses the generalization abilities enabling them to generate control pulses for the systems with variable strength of the disturbance. The presented results suggest that in some quantum control scenarios, geometrical data representation and processing is competitive to more complex methods.
\end{abstract}

\maketitle


\section{Introduction}

Spectacular progress made during the last few years rises hopes for the near-term 
construction of the devices based on the principles of quantum information 
processing~\cite{Riedel2017, Acin2018}. These principles can be used in 
metrology to develop new sensors and clocks~\cite{Degen2017}, to advance the 
abilities for simulating materials or chemical reactions~\cite{Johnson2014}, to 
improve the security of data transmission~\cite{Benatti2010}, and to extend the 
capabilities of computing equipment~\cite{Hirvensalo2004, Soeken2018}. One of 
the enabling ingredients for developing quantum technologies is the physical 
implementation of the basic ingredients of the quantum computational process, 
namely quantum gates. The development of the methods for systematic driving the 
quantum dynamical system from an initial state to the desired state is the 
subject of quantum optimal control~\cite{dalessandro2007introduction, Brif2010, 
   	Glaser2015}. Advancements in this area are crucial for the real-world applications 
of quantum information processing.

The problem of finding the sequence of physical operations implementing the 
given quantum logical operation has been studied in many 
scenarios~\cite{heule2010local,Pawela2016,PhysRevLett.106.190501,khaneja2007shortest,PhysRevLett.114.200502}.
As the problem can be very demanding numerically, a significant research 
effort has been invested in creating 
algorithms~\cite{GRAPE,CRAB,oleg2018krotov} and developing dedicated software 
packages~\cite{DYNAMO,qutip2}, focused on the optimization problems arising in 
this area.

As the main research focus in the area of developing new methods for quantum 
control is on the gradient-based methods, recently the utilization of machine 
learning techniques has been proposed to address these types of optimization 
problems. In particular, in \cite{Wu2016} a learning-based open-loop method to 
find robust control pulses for generating a set of universal quantum gates was 
proposed. In \cite{Kuang2018} a sampling-based approximate time-optimal control 
algorithm for a system with Hamiltonian fluctuations was discussed. The 
utilization of reinforcement learning in quantum control has also been 
proposed~\cite{DaoyiDong2008,ChunlinChen2013,fosel2018reinforcement,vedaie2018reinforcement,bukov2018reinforcement,niu2019universal}.
In~\cite{augustlehrstuhl} the connection 
between tensor networks and machine learning in the context of optimization has 
been presented (see also \cite{august2017using,august2018taking}).
Moreover, recently it has been demonstrated that machine learning is able to 
perceive the parameters of quantum systems \cite{iten2018discovering}. This 
suggests that the utilization of machine learning, especially in the form of 
artificial neural networks, can be beneficial for purpose of quantum 
control~\cite{Dunjko_2018}. However, little attention has been devoted to the
utilization of the simple machine learning techniques. 

The aim of our work is to explore the possibilities of utilizing machine 
learning techniques in the analysis of the mapping between control pulses
of an idealized system and control pulses of a system with drift. The 
approximation of this mapping is relevant for the understanding of the 
manifold of control pulses in noisy quantum systems. 
In~\cite{ostaszewski18approximation} we suggested that recurrent neural 
networks can be used as a method for constructing such approximation.
In this paper we scrutinize this approach in the context of classical machine 
learning methods. We compare recurrent neural networks with geometrical 
methods, and study the features and limitations of both approximation techniques.
In the first approach, the pulses are treated as time-series.
For harnessing this representation we use an artificial neural network designed
for time-series as the corresponding method~\cite{hochreiter1997long}. In the 
second approach, we treat the control pulse vectors as points in space, 
without assigning any meaning to the dimensions of the space. In this case, we 
use k-means clustering and \kNN classifier as the methods based on geometrical 
relations~\cite{kNN}.

In particular, there are two main observations that we support with the 
conducted numerical experiments.
\begin{itemize}
   	\item There exist nontrivial quantum control scenarios for which 
   	a simple, geometrical data representation and processing is competitive 
   	to more complex methods.
   	\item Interpolation and extrapolation with respect to the quantum system parameters requires a sophisticated machine learning model.
\end{itemize}
For that purpose we define two learning tasks based on one underlying quantum
control problem. These tasks allow us to highlight the similarities and
differences in the studied methods

We focus on the problem of controlling the closed quantum 
system~\cite{breuer2002theory}. Our goal is to learn the type of the 
disturbance in the system and subsequently to utilize this knowledge for 
generating control pulses for the system with the same kind of disturbance but 
acting with a different strength. We achieve this by encoding the information 
about the turbulence (drift) present in the system using the machine learning 
methods and exploiting their generalization abilities.

Finally, in this work, in contrast to the common 
approach, we utilise random unitary matrices and because of this  we probe the 
full space of unitary gates. The presented approach opens new possibilities for the utilization of machine learning techniques in the area of quantum control.

The rest of this paper is organized as follows. In 
Section~\ref{sec:preliminaries} we introduce the theoretical model of the quantum 
system considered in our study. 
Section~\ref{sec:results} contains the results of the numerical  experiments comparing the considered machine learning approaches to quantum control pulses. We introduce 
two setups which are subsequently used to benchmark the efficiency 
of both approaches to the representation of quantum control pulses. Final 
remarks and conclusions are presented in Section~\ref{sec:fianl}. Technical 
details of the machine learning methods used in the paper are provided in the 
Appendices. 

\section{Preliminaries -- description of quantum dynamcis}
\label{sec:preliminaries}

Let us consider a two-dimensional quantum system, a qubit, described by a state 
vector $\ket{\psi}\in\Cplx^2$. The evolution of the system is described by 
Schr\"odinger equation of the form
\begin{equation}
\frac{\dd\ket{\psi}}{\dd t} =-\ii H(t)\ket{\psi},
\label{eq:schrod}
\end{equation}
where $H$ is the Hamiltonian operator, representing the energy of the system.
In our model, the Hamiltonian $H$ is a sum of two terms
\begin{equation}
H(t) = H_c(t) + H_{\gamma},
\end{equation}
corresponding to the
\emph{control field} and the \emph{drift} interaction, respectively.
The control Hamiltonian $H_c$
\begin{align}
H_c(t) &= h_x(t)\Sx+ h_z(t)\Sz, \label{eq:control_hamiltonian} 
\end{align}
with $\Sx=\left(\begin{smallmatrix}0&1\\1&0\end{smallmatrix}\right)$, 
$\Sz=\left(\begin{smallmatrix}1&0\\0&-1\end{smallmatrix}\right)$,
represents the external fields used to alter the dynamics of the system.

On the other hand, the drift Hamiltonian $H_{\gamma}=\gamma \Sy$, where 
$\gamma$ is a real parameter and  $\Sy=\ii\Sx \Sz$, represents the 
intrinsic dynamics of the system which is independent of the external 
interactions. Parameter $\gamma$ describes the strength of this dynamics and  is treated as a disturbance in 
our physical model.

To steer the system, one needs to choose the coefficients in 
Eq.~\eqref{eq:control_hamiltonian}, \ie $h(t) 
= (h_x(t),h_z(t))$. In this model, we assume that 
function $h(t)$ is constant in time intervals $\Delta t_i = 
[t_i,t_{i+1})$, which are the elements of the equal partition of the evolution time interval $T 
=\bigcup_{i=0}^{n-1}\Delta t_i$. Thus, control parameters $h(t)$ will be denoted as a vector of values for time 
intervals. Moreover, we assume that $h(t)$ has values 
from the interval $[ -1,1]$. 
Solving Eq.~\eqref{eq:schrod} we obtain the equation on motion for the system
\begin{equation}
\ket{\psi_T}=U(h(T))\ket{\psi},
\end{equation}
where
\begin{equation}
U(h(T))=e^{ H(\Delta t_n)}\cdots e^{ H(\Delta t_2)}e^{ H(\Delta t_1)}
\label{eq:integrate_superoperator}
\end{equation}
is the unitary matrix governing the time-evolution of the state vector. In the following, we will skip $T$ in the above formula, i.e. we will note $U(h)$. 
The central problem of quantum control is to choose the appropriate control 
pules, $h_x$, $h_z$, given the requested resulting unitary matrix.

The figure of merit in this problem is the \emph{fidelity distance between 
  superoperators}, defined as \cite{floether12robust}

\begin{equation}
F=1-F_{err},
\end{equation} 
with
\begin{equation}
F_{err} = \frac{1}{2N^2}  \tr \left[ (A-B(h))^\dagger (A-B(h))\right] ,
\label{eq:fidelity-error}
\end{equation} 
where $N$ is the dimension of the system in question, $A$ is the superoperator of 
the fixed target operator,
and $B(T)$ is the evolution superoperator of operator resulting from the numerical 
integration of Eq.~\eqref{eq:schrod}
with given controls.  In particular, for a target unitary operator $U$, its 
superoperator 
$A$ is given by the formula
\begin{equation}
A=U\otimes \bar{U},
\end{equation}
where $\bar{U}$ is the element-wise conjugate of $U$.
Superoperator $B(h)=U(h)\otimes \bar{U}(h)$ is obtained from the unitary operator resulting from the 
integration of the Eq.~(\ref{eq:schrod}).

\section{Assessment of machine learning abilities}\label{sec:results}

\subsection{Formulation of the problem}

We assume that we have a qubit quantum system with two control fields as 
described by Eq.~\eqref{eq:control_hamiltonian}. Moreover, we are given a 
set of unitary operators and corresponding control signals $h(t)$ that can be implemented in the physical system. In our consideration, we 
introduce a drift $H_\gamma$ to the system -- an additional constant  
term in the system Hamiltonian. In the presence of the drift, the control  
signals calculated for the system without the drift cannot be used to obtain  
the desired evolution. One way to fix this is to estimate the drift and then  
recalculate all of the control signals. The drawback of this approach is that  
it requires the explicit knowledge of the drift Hamiltonian.

In the proposed approach, we use a machine learning method to solve the issue 
without learning the drift explicitly. Namely, the analysed machine learning  
task is used to learn a general method to modify the signals so that they would
work with the system with the drift.

\subsection{Utilization of machine learning}

In order to study the utilization of machine learning for solving the above
problem, we introduce two tasks centred around the concept of quantum control
and approach both of them with machine learning methods. In both tasks, we use
different methods of data representation, and each task is used to build a test
environment for comparing the selected methods.

Both tasks are based on the same scenario and in both cases, the goal is to
counteract an unknown drift occurring in the controlled quantum system.  We use
machine learning (ML) algorithms to \emph{translate} the control pulses,
$\NCP$, calculated for the system without drift, into the control pulses
counteracting the drift, $\DCP$. General scheme for considering both tasks is
presented in Fig.~\ref{fig:task-overview}. The ability to successfully translate
between control pulses is equivalent to the ability to learn the structure of
the unknown drift. Thus, the approximations based on machine learning should map
control pulses $\NCP$, into control pulses, $\nnDCP \approx \DCP$, which should be effective
for controlling the system with the unknown drift. One can note that the mapping from \NCP to \DCP shares the mathematical properties 
with the problem of statistical machine translation \cite{koehn2009statistical}, a problem which is 
successfully modelled with artificial neural networks~\cite{bahdanau2014neural}. 

We compare the effectiveness
of ML by comparing the efficiency in terms of fidelity error and the generalization abilities.  By the efficiency we understand the mean fidelity error, taken over target unitary matrices, between the operators generated from \DCP using the considered approximation and target operators.

\begin{figure}[ht!]
    \centering
    \subfigure[\ Quantum control in ideal system.]{
    \begin{tikzpicture}
    [>=latex,
    ->,shorten >=1pt,
    cell/.style={
        align=center,
        rectangle, 
        rounded corners=2mm, 
        draw
    },
    input/.style={
    },
    hidden/.style={
    },
    subcell/.style={
        align=center,
        rectangle, 
        draw,
        minimum width = 1.cm,
        minimum height = 0.7cm,
        rounded corners=1mm
    },
    ]

    \node[minimum size=20pt](h1) at (0,-1) {\NCP};
    \node[draw,subcell,minimum size=30pt](o1) at (2,-1) {QS($\gamma=0$)};
    \node[minimum size=20pt](o2) at (4,-1) {$U$};

    \draw [->,>=stealth] (h1) --node[above]{}  (o1);
    \draw [->,>=stealth] (o1) --node[above,pos=0.8]{}  (o2);
    \node[above,font=\large\bfseries] at (current bounding box.north) {};
    
    \end{tikzpicture}

    \label{fig:execution_ncp}}
    \centering
\subfigure[\ Quantum control in the system with drift.]{
    \begin{tikzpicture}
    [>=latex,
    ->,shorten >=1pt,
    cell/.style={
        align=center,
        rectangle, 
        rounded corners=2mm, 
        draw
    },
    input/.style={
    },
    hidden/.style={
    },
    subcell/.style={
        align=center,
        rectangle, 
        draw,
        minimum width = 1.cm,
        minimum height = 0.7cm,
        rounded corners=1mm
    },
    ]
    
    \node[minimum size=20pt](h1) at (0,-1) {\DCP};
    \node[draw,subcell,minimum size=30pt](o1) at (2,-1) {QS($\gamma>0$)};
    \node[minimum size=20pt](o2) at (4,-1) {$U$};
    
    \draw [->,>=stealth] (h1) --node[above]{}  (o1);
    \draw [->,>=stealth] (o1) --node[above,pos=0.8]{}  (o2);
    \node[above,font=\large\bfseries] at (current bounding box.north) {};
    
    \end{tikzpicture}\label{fig:execution_dcp}}
    \centering
    \subfigure[\ Correction of quantum control using machine learning.]{
        \begin{tikzpicture}
        [>=latex,
        ->,shorten >=1pt,
        cell/.style={
            align=center,
            rectangle, 
            rounded corners=2mm, 
            draw
        },
        input/.style={
        },
        hidden/.style={
        },
        subcell/.style={
            align=center,
            rectangle, 
            draw,
            minimum width = 1.cm,
            minimum height = 0.7cm,
            rounded corners=1mm
        },
        ]

        \node[minimum size=20pt](h1) at (2,-1) {\NCP};
        \node[draw,subcell,minimum size=30pt](ML) at (4,-1) {ML};
        \node[minimum size=20pt](o1) at (6,-1) {\nnDCP};
        \draw [->,>=stealth] (h1) --node[above]{}  (ML);
        \draw [->,>=stealth] (ML) --node[above,pos=0.8]{}  (o1);
        \node[above,font=\large\bfseries] at (current bounding box.north) {};
        
        \end{tikzpicture}\label{fig:correction_scheme}
    }  
    
    \caption{Learning the structure of the drift. Here $\NCP$ represents control
    pulses implementing unitary matrix $U$ in the quantum system QS without the drift, 
    $\DCP$ respective control pulses implementing this matrix in the 
    system with the additional drift, and $\nnDCP$ control pulses generated
    using one of the machine learning methods. Our goal is to learn the method of 
    transforming from \NCP to \nnDCP. This enables us to utilize the obtained ML
    algorithm for generating control pules for the system with the drift, even
    in the situation where the strength of the drift was not available during
    the learning phase.}
    \label{fig:task-overview}
\end{figure}

The unknown drift occurring in the studied quantum system is controlled by the
strength $\gamma$. In our study, we utilize this parameter to probe ML methods.
We do this by changing the way of altering this parameter in both tasks.

In the first task, our goal is to asses the efficiency of ML methods. For this
reason, the drift strength is fixed during the training and testing phases. ML
algorithm utilizes pairs of corresponding $\NCP$ and $\DCP$ pulses to learn
about the structure of the drift. During the testing phase, we check if the
$\nnDCP$ pulses can be used to control unitary evolution with high fidelity, and
check which ML methods provide best results.  

In the second task, the drift strength is variable. We consider a few values of 
the drift strength during the training phase and an extended values set during the test 
phase. This enables us, apart from testing the efficiency, to test the generalization 
abilities of the considered methods.

One should note that control pulses $\NCP$ can be seen as vectors of pairs, with
the first dimension corresponding to the time slot $\{t_0, \ldots, t_{n-1}\}$,
and the second dimension corresponding to the values of control fields
$\{h_x,h_z\}$,
\begin{equation}
\begin{split}
\NCP_{i,j} &= h_j(t_i),
\end{split}
\end{equation}
with $i \in \{0,\ldots ,n-1\}$, $j\in\{x,z\}$. Hence, it is equally natural to treat
these pulses as geometrical objects, as well as time-series. In this first case, the
natural machine learning method is given by the algorithms analysing the distance
between vectors, whilst if we want to harness the time-series structure it is natural
to employ artificial neural networks.

\subsection{Task I: Efficiency with constant drift}

In the first task, we deal with constant drift strength. Our goal is to check 
the efficiency of machine learning measured by the fidelity between the unitary matrix implemented
using and $\DCP$, and the unitary matrix obtained by using pulses $\nnDCP$, generated using machine
learning. The considered scenario is presented in Fig.~\ref{fig:ML_experiment_scheme}.

\begin{figure}[h!]
    \centering
    \subfigure[\ Training phase.]{
        \begin{tikzpicture}
        [>=latex,
        ->,shorten >=1pt,
        cell/.style={
            align=center,
            rectangle, 
            rounded corners=2mm, 
            draw
        },
        input/.style={
        },
        hidden/.style={
        },
        subcell/.style={
            align=center,
            rectangle, 
            draw,
            minimum width = 1.cm,
            minimum height = 0.7cm,
            rounded corners=1mm
        },
        ]
        
        \node[minimum size=38](in1) at (-.5,-1) {$U$}; 
        \node[minimum size=38](in2) at (-.5,-3) {$U$}; 
        
        \node[minimum size=20pt](h1) at (2,-1) {\NCP};
        \node[minimum size=20pt](h2) at (2,-3) {\DCP};
        
        \node[draw,subcell,minimum size=40pt](ML) at (4,-2) {ML};
        
        \draw [->,>=stealth] (in1) --node[above] {GRAPE} (h1);
        \draw [->,>=stealth] (in1) --node[below] {$\gamma=0$} (h1);
        \draw [->,>=stealth] (in2) --node[above] {GRAPE} (h2);
        \draw [->,>=stealth] (in2) --node[below] {$\gamma>0$} (h2);
        
        \draw [->,>=stealth] (h1) --node[above]{}  (ML);
        \draw [->,>=stealth] (h2) --node[above,pos=0.8]{}  (ML);
        \node[above,font=\large\bfseries] at (current bounding box.north) {};
        
        \end{tikzpicture}\label{fig:training_scheme}}
    \centering
    \subfigure[\ Testing phase.]{
        \begin{tikzpicture}
        [>=latex,
        ->,shorten >=1pt,
        cell/.style={
            align=center,
            rectangle, 
            rounded corners=2mm, 
            draw
        },
        input/.style={
        },
        hidden/.style={
        },
        subcell/.style={
            align=center,
            rectangle, 
            draw,
            minimum width = 1.cm,
            minimum height = 0.7cm,
            rounded corners=1mm
        },
        arrow/.style={
            rounded corners=.2cm
        }
        ]
        
        \node[minimum size=38](in1) at (-1,-1) {$U$}; 
        \node[minimum size=20pt](h1) at (2,-1) {\NCP};
        \node[draw,subcell,minimum size=40pt](ML) at (4,-1) {ML};
        \node[minimum size=20pt](o1) at (6,-1) {\nnDCP};
        \node[minimum size=20pt](o2) at (9,-1) {$U(\nnDCP)$};
        \node[minimum size=20pt](o3) at (12,-1) {$F_{err}(U(\nnDCP), U)$};
        
        \draw [->,>=stealth] (in1) --node[above] {\qutip} (h1);
        \draw [->,>=stealth] (in1) --node[below] {$\gamma=0$} (h1);
        
        \draw [->,>=stealth] (h1) --node[above]{}  (ML);
        \draw [->,>=stealth] (ML) --node[above,pos=0.8]{}  (o1);
        \draw [->,>=stealth] (o1) --node[above]{QS($\gamma>0$)}  (o2);
        \draw [->,>=stealth] (o2) --node[above]{}  (o3);
        \node[above,font=\large\bfseries] at (current bounding box.north) {};
        
        \end{tikzpicture}\label{fig:test_scheme}
    }
    \caption{Task I: Assessment of the machine learning efficiency.
        We assume that we need to implement unitary matrix $U$ by controlling Quantum System (QS).
        \subref{fig:training_scheme} Training phase. We use \qutip package to generate control pulses \NCP and \DCP for $\gamma=0$ and $\gamma>0$
        respectively. In the training phase, pairs (\NCP, \DCP) are used to train the machine learning
        algorithm. \subref{fig:test_scheme} Testing phase. In the testing phase, machine learning algorithm is used to translate \NCP into 
        control pulses \nnDCP which implements target unitary matrix $U_k$ on a quantum system with undesired
        drift, QS($\gamma>0$).}
    \label{fig:ML_experiment_scheme}
\end{figure}

The procedure describing numerical experiments executed in the scope of Task I 
consists of the following steps.
\begin{enumerate}
  \item \textbf{Generating training set}
  Generate samples of Haar random 
  unitary matrices~\cite{mezzadrigenerate,miszczak12generating}, $U_k$, and using GRAPE algorithm implemented in \qutip~\cite{qutip,qutip1,qutip2}, 
  generate (\NCP, \DCP) pairs corresponding to $U_k$ and fixed $\gamma>0$.
  
  In the presented numerical experiments, GRAPE algorithm is always initialized by identity matrix and the exact implementation of sampling random Haar unitary matrices 
  is available at~\cite{qcontrol_lstm_approx}. All numerical experiments are performed for evolution time $T=2.1$, divided into sixteen intervals $n=16$. 
  The range of values of the parameter $\gamma$ is restricted to the interval $[0,1]$. This is due to the fact that it is adjusted to the restrictions for the parameters of the physical model.
  One should note that we were not able to find satisfactory control pulses for higher values of $\gamma$ using the gradient-based method.
  
  \item \textbf{Training} To create correction scheme (see Fig.~\ref{fig:correction_scheme}) we perform the following steps.
  \begin{enumerate}
    \item \textbf{Train artificial neural network} Utilize \NCP as an input and treat the corresponding \DCP as an output reference. As a loss function, we utilize mean squared error between target \DCP, and prediction of artificial neural network \nnDCP. The input of the LSTM is a pair $x_t=(\NCP_x(t),\NCP_z(t))$ (see Appendix~\ref{sec:LSTM}).
    
    \item \textbf{Train the clustering correction algorithm}
    Using training pairs (\NCP, \DCP) compute corrections, called correction control pulses, $\CCP = {\DCP}-{\NCP}$. Next, evaluate $k$-means algorithm on the $\CCP$. Finally, compute average correction control pulse $\tilde{C}_j$, for each cluster resulting from $k$-means algorithm, where $j$ is an index of a cluster.
    Detailed description of the clustering correction algorithm is provided in Appendix~\ref{sec:clusterring_corr}.
    
    It it important to note that by clustering \CCP pulses, we obtain natural clusters on \NCP pulses. Therefore, we can execute \kNN algorithm on test $\NCP$ and the training set to determine the correction, which is equivalent to finding the desired $\nnDCP$. The \kNN algorithm has number of neighbours set to $k=4$.
   
  \end{enumerate}

  \item \textbf{Testing}
  The procedure consists of the following steps (see Fig.~\ref{fig:test_scheme}). 
  \begin{enumerate}
    \item Generate Haar random unitary matrices $U_{k}$.
    
    \item Use GRAPE to generate a testing set consisting of \NCP for $U_{k}$.
    
    \item Use the trained ML algorithm to generate $\nnDCP$ for each \NCP in the testing set.
    
    \item Use $\nnDCP$ to construct operator $U(\nnDCP)$ assuming the drift $\gamma>0$.
    
    \item Use $\NCP$ to construct operator $U(\NCP)$ from testing \NCP, also
    assuming the drift $\gamma>0$. This provides a reference point, \ie the difference between $F_{err}(U_k,U(\NCP))$ and $F_{err}(U_k,U(\nnDCP))$, reflecting the quality of the correction \nnDCP of \NCP.

  \end{enumerate}
\end{enumerate}

It should be noted that the algorithms used in the numerical experiments directly work on (\NCP, \DCP) pairs. However, the evaluation of the efficiency requires the comparison on the level of operators corresponding to the considered control pulses. Therefore, (\NCP, $U_{k}$) pairs are used in the testing data. The efficiency is calculated as the \emph{mean fidelity} over the set of unitary matrices $U_k$,
\begin{equation}
\frac{1}{M}\sum_{k=1}^M (1- F_{err}(U_k, U(h_k))),
\end{equation}
where $h_k$  are control pulses corresponding to unitary matrix $U_k$ and $M$ is the number of unitary matrices. For the sake of clarity, we do not use the indexing of the control pulses if the relation between them and the unitary matrix can be deduced from the context.

We can distinguish two reference values for which the efficiency of the considered approximation is compared, namely 

\begin{itemize}
    \item[(R1)] the mean fidelity between operators obtained from \NCP, applied on a system with drift and target operators, obtained in the last step of testing,
    \item[(R2)] one \ie the maximal value which can be obtained from fidelity function.
\end{itemize}

It is reasonable to obtain the efficiency higher than the value from point~(R1), which represents the situation of not using any correction at all. However, the closer the efficiency of the methods to the value from point~(R2), the higher its effectiveness in the described scenario.

\subsubsection{Efficiency of LSTM}
As it has already been mentioned, the sequence of control pulses can be naturally represented as a time series. This type of data is in many cases analysed using a specialised type of artificial neural networks, namely Long Short-Term Memory (LSTM) neural networks, developed harnessing time-series character of the data. Thus, we can use LSTM as the 
approximation function translating the correction scheme for control pulses. 
The trained artificial neural network will be used as a map from \NCP to \nnDCP.


In numerical experiments executed to asses the efficiency of LSTM, we generate 5000 random unitary matrices and train the neural network to predict \DCP based on \NCP. For each value of gamma, a new artificial neural network is trained. The network is given the reference (\NCP, \DCP) pairs as a training set. For each $\gamma$ value, the following results were obtained with LSTM networks trained on 3000 control pulses. 
The tests were performed on the set of 2000 control pulses.  

As one can see in Table~\ref{tab:results-lstm}, the results obtained by the GRAPE algorithm are better than those obtained from LSTM network. This effect is caused by the fact that the artificial neural network utilizes the data generated from gradient-based method as target data. However, the results presented in Table~\ref{tab:results-lstm} demonstrate that LSTM network can achieve efficiency with errors of the same order of magnitude as the reference data. Moreover, one should note that for higher values of gamma parameter the data obtained from GRAPE contain many outliers, \ie there is a significant number of matrices for which the resulting fidelity is below the acceptable level of 0.90.

\subsubsection{Efficiency of clustering}
The second ML method we use is based on the purely geometrical representation of the data. To exploit the geometrical representation of the control pulses, we use $k$-means algorithm to utilize mean control from the cluster to correct \NCP.  From the computational point of view, the advantage of this method is its simplicity. One should also note that the geometrical representation does not assign any order to the coordinates of the pulses and relies on the distance between vectors only.

As the input data, we take 3000 random unitary matrices and generate corresponding \DCP, \NCP and \CCP vectors. Here our goal is to check if the resulting clustering provides efficient correction. To this end, for each cluster, we calculate mean correction and apply it to \NCP from respective clusters. Next, we calculate the mean fidelity for operators obtained from
\nnDCP signals and compare it with the efficiency of \DCP and the efficiency of \NCP applied on the system with undesired drift. Moreover, we repeat this method on the approximations of \CCP presented in section~\ref{sec:approx_fncs}.


\begin{figure}[ht!]
  
  \subfigure[Case $\gamma=0.2$, where average fidelity of \NCP is .90]{
    \includegraphics[width=0.45\textwidth]{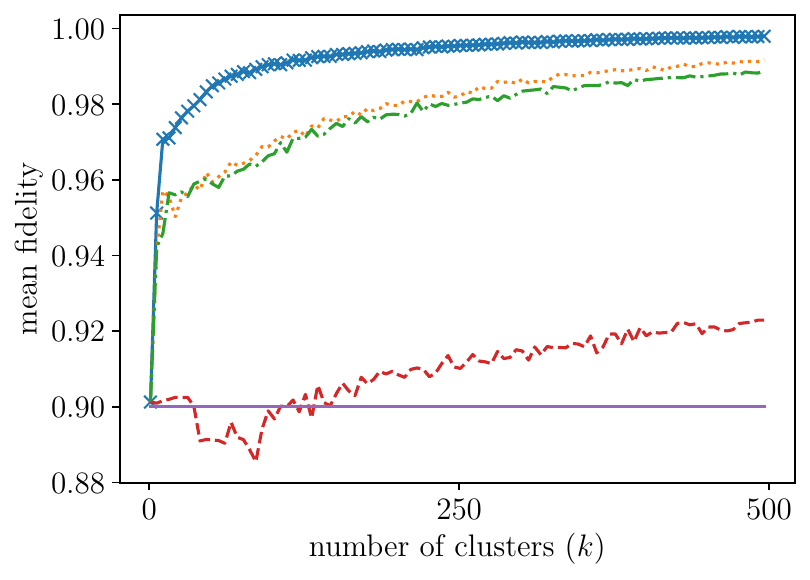}
    \label{fig:eff_vs_nb_of_clusters_gam_02}}
  \subfigure[Case $\gamma=0.4$, where average fidelity of \NCP is .66]{
    \includegraphics[width=0.45\textwidth]{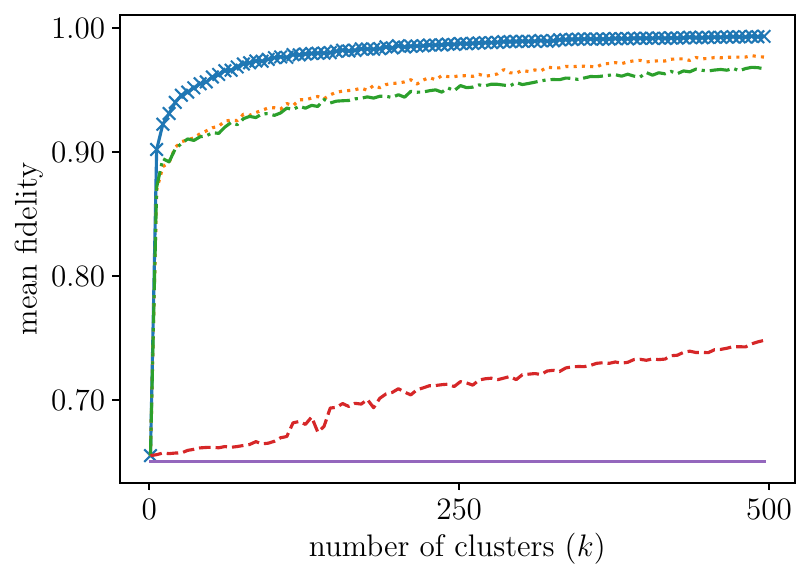}
    \label{fig:eff_vs_nb_of_clusters_gam_04}}
  \subfigure[Case $\gamma=0.6$, where average fidelity of \NCP is .37]{
    \includegraphics[width=0.45\textwidth]{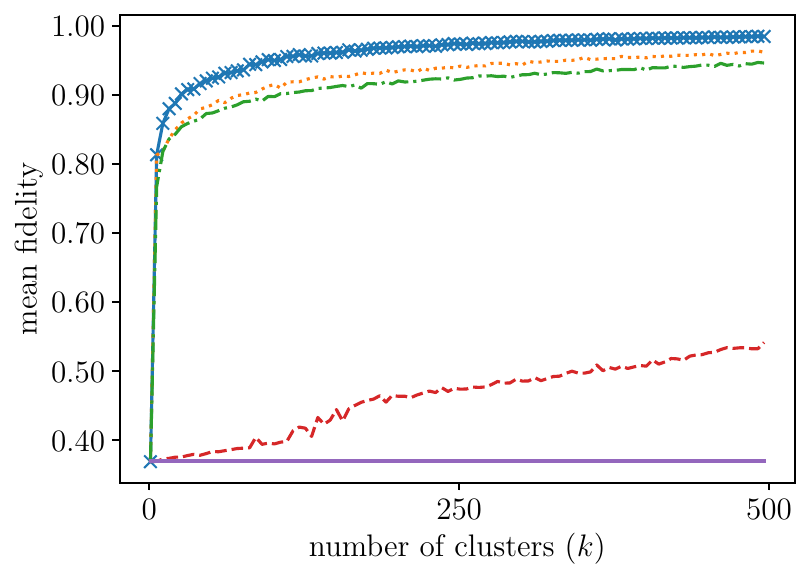}
    \label{fig:eff_vs_nb_of_clusters_gam_06}}
  \subfigure[Case $\gamma=0.8$ , where average fidelity of \NCP is .16]{
    \includegraphics[width=0.45\textwidth]{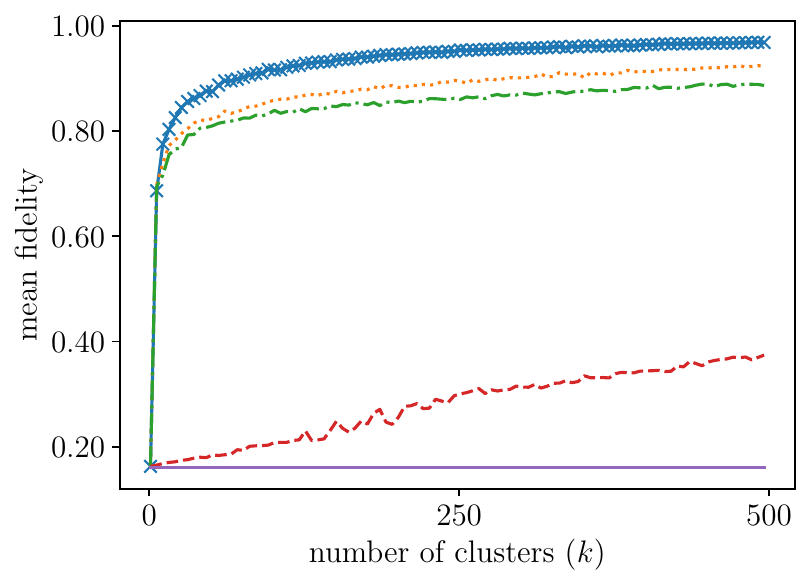}
    \label{fig:eff_vs_nb_of_clusters_gam_08}}
  \caption{The comparisons of  the efficiency of corrections from clustering 
    with the efficiency of NPC. The analysis was performed on 3000 samples. 
    The 
    number of clusters on axis $x$ varies from 1 to 496 with step 5.
    Values marked with blue $\times$ correspond to clustering directly on 
    \CCP. Orange dotted line corresponds to clustering on approximated data 
    by 
    $f_{poly4}$ function, and dot-dash green line corresponds to clustering 
    on approximated data by 
    $f_{poly3}$ (see Appendix~\ref{sec:approx_fncs}).
    Red dashed lines corresponds to the clustering on approximated data by 
    combination of sinusoid.        
    Purple continuous line corresponds to the average fidelity 
    obtained from the application of \NCP on the system with drift 
    Hamiltonian.}
  \label{fig:comparison_approx_3000}
\end{figure}

As one can see in Fig.~\ref{fig:comparison_approx_3000}, for each kind of
approximation method, there exists $k$ such that applying the correction 
scheme, being mean \CCP within corresponding cluster, is better than the mean 
fidelity of application of \NCP with $\gamma\neq 0$. Moreover, 
the clustering on the raw data, without any kind of approximation, yields 
significantly better results in comparison to approximation variants. This demonstrates that the methods used for 
approximation are not suitable for reducing the dimensionality in this case.

\begin{figure}[h!]
  \centering
  \includegraphics[width=0.5\textwidth]{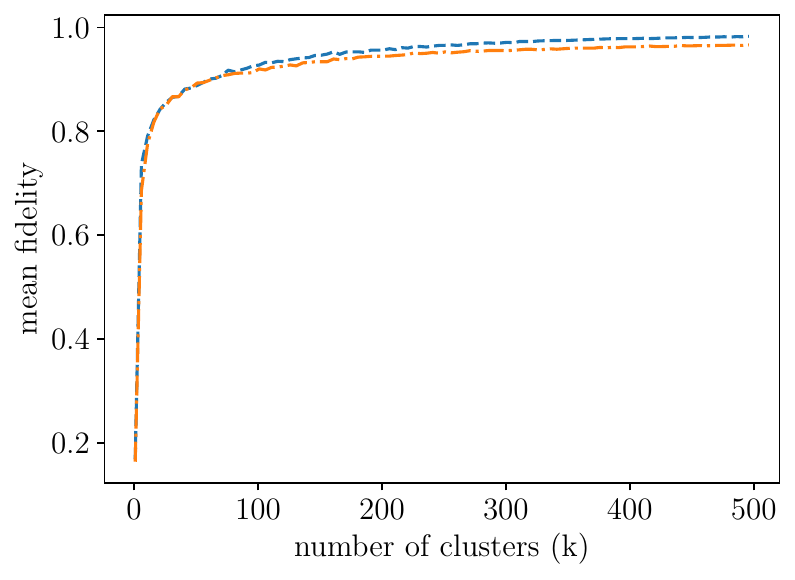}
  \caption{Insensitivity of the clustering to the size of the training set. The graph plotted with a blue dashed line corresponds to the 
    sample of size 1000, and the graph plotted with a orange dot-dashed line 
    corresponds to the sample of size 5000. For number of clusters equal to 500, we saturate the abilities of the clustering. In the first case the average cluster contains 2 elements, whilst in the former it contains 10 elements.
    The numerical experiment was performed for $\gamma = 0.8$.}
  \label{fig:comparison_1000_vs_5000_samples}
\end{figure}

In Fig.~\ref{fig:comparison_1000_vs_5000_samples} one can see the important 
characteristics of the algorithm, namely its insensitivity to the 
number of samples in the training set. One can see that the clustering with a fixed parameter $k$ gives similar results for training sets consisting of 1000 and 5000 samples. The obtained results suggest that clustering can be used to achieve high fidelity of the control pulses even for the situation where only a small number of examples is used.

\begin{figure}[ht!]
  \centering
  \subfigure[Case $\gamma=0.2$, where average fidelity of \NCP is .90]{
    \includegraphics[width=0.45\textwidth]{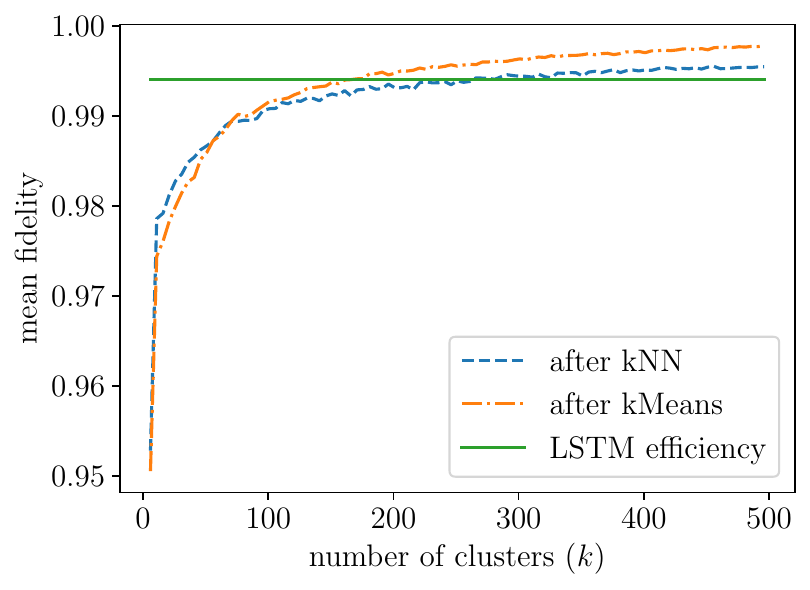}
    \label{fig:kNNeff_vs_nb_of_clusters_gam_02}}\quad
  \subfigure[Case $\gamma=0.4$, where average fidelity of \NCP is.65]{
    \includegraphics[width=0.45\textwidth]{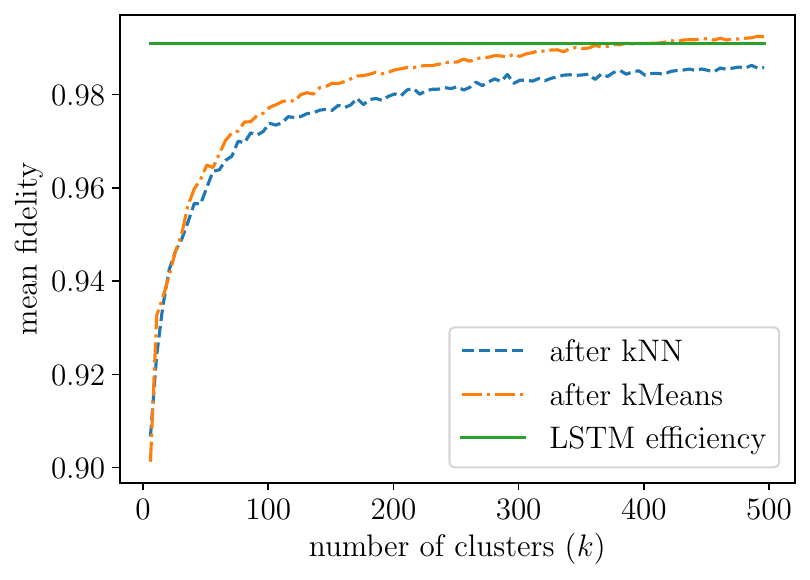}
    \label{fig:kNNeff_vs_nb_of_clusters_gam_04}}
  \subfigure[Case $\gamma=0.6$, where average fidelity of \NCP is .37]{
    \includegraphics[width=0.45\textwidth]{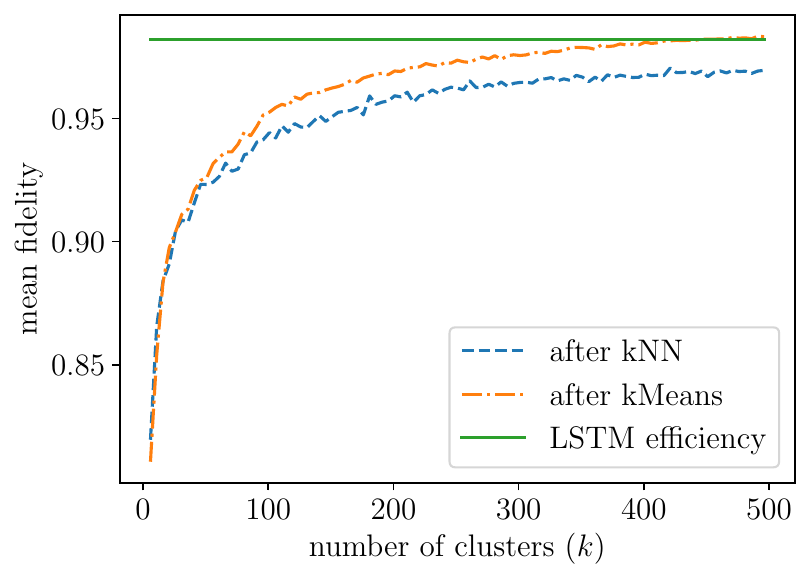}
    \label{fig:kNNeff_vs_nb_of_clusters_gam_06}}
  \subfigure[Case $\gamma=0.8$, where average fidelity of \NCP is .16]{
    \includegraphics[width=0.45\textwidth]{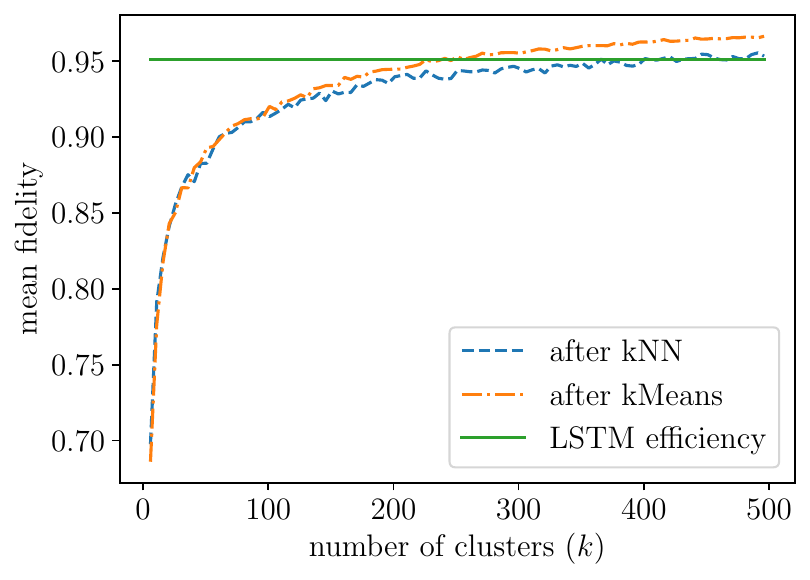}
    \label{fig:kNNeff_vs_nb_of_clusters_gam_08}}
  \caption{The above plots show the comparisons of the efficiency of 
    clustering 
    on a training set (3000 samples) and classifier on a test set (2000 
    samples) 
    with respect to the number of clusters, \ie the number of universal 
    correction 
    schemes. Because of low divergence, we choose one representative $k$ in 
    kNN 
    algorithm \ie $k=4$.}
  \label{fig:kNN_efficiency}
\end{figure}

The last part of the assessment procedure consists of the efficiency test of the
whole correction scheme, \ie we aim at answering what is the efficiency when we use 
kNN classifier (see Sec.~\ref{sec:kNN}).
In this case the efficiency of \kNN should be limited by the efficiency resulting from clustering.
As we can see in Fig.~\ref{fig:kNN_efficiency}, \kNN classifier has
similar results on the test set as the clustering on the training set. 

\subsubsection{Comparison of the efficiency}

The main conclusion one can draw from the obtained results is that the clustering methods can achieve surprisingly good efficiency in the considered task. One can see in Fig.~\ref{fig:kNN_efficiency} that for the number of clusters close to 300, we obtain the efficiency similar to LSTM.

This behaviour suggests that the quantum control pulses have a geometrical structure which can be captured and harnessed using $k$-means and kNN algorithms. The results obtained using this approach are very similar to the results obtained using a significantly more complicated approach represented by LSTM. This suggests that for the considered task, the utilization of artificial neural networks is not the only machine learning method which should be considered.

\begin{table}[h!]
  \centering
  \begin{tabular}{ l@{\hspace{0.42cm}}c c c c}\hline &
    \multicolumn{4}{c}{$\gamma$} \\\cline{2-5}
    Method  & $0.2$ & $0.4$ & $0.6$ & $0.8$ \\\hline
    \nnDCP from LSTM			& $.994$ & $.991$ & $.982$ & $.951$  \\
    \DCP from GRAPE			& $.999$ & $.999$ & $.998$ & $.989$  \\
    \NCP from GRAPE			& $.903$ & $.657$ & $.372$ & $.165$
  \end{tabular}
  \caption{Comparison of efficiency (mean fidelity) for $H_{\gamma}=\gamma\Sy$ for 
    different methods of obtaining control pulses. Time of the evolution was 
    set to $T=2.1$. We restricted the magnitude of control pulses to interval $[-1, 1]$. The number of time slots was set to $n=16$. Fidelity obtained using the artificial neural network  (\nnDCP) is very similar to the efficiency 
    obtained using gradient method. As expected, fidelity obtained using the 
    orginal pulses (\NCP) deteriorates with the increasing strength of the 
    drift. 
  }
  \label{tab:results-lstm}
\end{table}

\subsection{Task II: Generalization with variable drift}

\subsubsection{Task Details}

Having assessed the efficiency of the machine learning methods in the scenario where the drift strength is fixed, we now analyse their generalization abilities. To this end we introduce the second task which extends the first by allowing the manipulation of the drift strength.

One should note that, as in Task I, we assume that the drift is unknown during the learning phase. However, in contrast to the situation in Task I, now we train out algorithm using different system parameters. In other words, the algorithm is given not only the information about the pairs $(\NCP, \DCP)$, but additionally each pair is associated with the drift strength. The task is to learn how to modify the control signals so that it would work for the values of $\gamma$ not included in the learning examples. The idea is illustrated in Fig.~\ref{fig:task-ii-overview}.

\begin{figure}[ht!]
  \centering
    \subfigure[\ Training phase.]{
    \begin{tikzpicture}
    [>=latex,
    ->,shorten >=1pt,
    cell/.style={
        align=center,
        rectangle, 
        rounded corners=2mm, 
        draw
    },
    input/.style={
    },
    hidden/.style={
    },
    subcell/.style={
        align=center,
        rectangle, 
        draw,
        minimum width = 1.cm,
        minimum height = 0.7cm,
        rounded corners=1mm
    },
    arrow/.style={
        rounded corners=.2cm
    }
    ]
    
    \node[minimum size=20](in1) at (-.5,-1) {$U$}; 
    \node[minimum size=20](in2) at (-.5,-3) {$U$}; 
    
    \node[minimum size=20pt](h1) at (2,-1) {\NCP};
    \node[minimum size=20pt](h2) at (2,-3) {(\DCPinA, $\gamma$)};
    
    \node[draw,subcell,minimum size=40pt](ML) at (4,-2) {ML};
    
    \draw [->,>=stealth] (in1) --node[above] {GRAPE} (h1);
    \draw [->,>=stealth] (in1) --node[below] {$\gamma=0$} (h1);
    \draw [->,>=stealth] (in2) --node[above] {GRAPE} (h2);
    \draw [->,>=stealth] (in2) --node[below] {$\gamma\in A$} (h2);
    
    \draw [->,>=stealth] (h1) --node[above]{}  (ML);
    \draw [->,>=stealth] (h2) --node[above,pos=0.8]{}  (ML);
    \node[above,font=\large\bfseries] at (current bounding box.north) {};
    
    \end{tikzpicture}

    \label{fig:execution_dcp_various_gamma}}
  
  \centering
    \subfigure[\ Testing phase.]{
    \begin{tikzpicture}
    [>=latex,
    ->,shorten >=1pt,
    cell/.style={
        align=center,
        rectangle, 
        rounded corners=2mm, 
        draw
    },
    input/.style={
    },
    hidden/.style={
    },
    subcell/.style={
        align=center,
        rectangle, 
        draw,
        minimum width = 1.cm,
        minimum height = 0.7cm,
        rounded corners=1mm
    },
    arrow/.style={
        rounded corners=.2cm
    }
    ]
    
    \node[minimum size=20](in1) at (-1,-0.5) {$U$}; 
   
    \node[minimum size=20pt](h1) at (1.8,-0.5) {$\NCP$};
    \node[minimum size=20pt](h2) at (1.8,0.5) {$ \gamma\not\in A$};
    
    \node[draw,subcell,minimum size=40pt](ML) at (4,0) {ML};
    \node[minimum size=20pt](o1) at (6,0) {\nnDCPniA};
    \node[minimum size=20pt](o2) at (9,0) {$U(\nnDCPniA)$};
    \node[minimum size=20pt](o3) at (12,0) {$F_{err}(U(\nnDCPniA), U)$};
    
    \draw [->,>=stealth] (in1) --node[above] {\qutip} (h1);
    \draw [->,>=stealth] (in1) --node[below] {$\gamma=0$} (h1);
    
    \draw [->,>=stealth] (h1) --node[above]{}  (ML);
    \draw [->,>=stealth] (h2) --node[above]{}  (ML);
    \draw [->,>=stealth] (ML) --node[above,pos=0.8]{}  (o1);
    \draw [->,>=stealth] (o1) --node[above]{QS($\gamma\not\in A$)}  (o2);
    \draw [->,>=stealth] (o2) --node[above]{}  (o3);
    \node[above,font=\large\bfseries] at (current bounding box.north) {};
    
    \end{tikzpicture}

    \label{fig:testing_diff_gamma}}
  \centering

  \caption{Task II: Learning correction for an unknown drift. 
      As in Task I, the goal is to implement unitary matrix $U$ in quantum system. The difference is in the role of the $\gamma$ parameter, which is not fixed during the task. \subref{fig:execution_dcp_various_gamma} Training phase. We use GRAPE algorithm from the \qutip package to generate control pulses \NCP and \DCPinA for selected values of $\gamma \in A$. \subref{fig:testing_diff_gamma} Testing phase. We use trained machine learning method to translate pairs $(\DCP, \gamma\not \in A)$ into \nnDCPniA for given $\gamma\in A$. Changes in the values of drift strength are used to model the changes of the internal dynamics of the system and enables the analysis of the generalization abilities.}\label{fig:task-ii-overview}
\end{figure}

During the testing phase, the trained algorithm also operates on the control pulses supplemented with the information about the target value of the drift strength. The values of parameter $\gamma$ occurring in this phase can differ from the values utilized in the learning phase. This scenario is presented in Fig.~\ref{fig:testing_diff_gamma}.

For the purpose of analysing the ability to generalize the correction scheme we have to fix the reference values, which will be used to benchmark the generalization efficiency of the utilized methods. For our purpose we define \emph{reference point of generalization} as the mean fidelity obtained by the algorithm trained on data with fixed parameter $\gamma$, applied to other values of the parameter. Such defined reference point provides the minimal efficiency which should be obtained by the tested algorithm in order to consider it suitable for using it for the new values of the drift strength.

The reference points are constructed as follows. Let us suppose that we have an already trained machine learning model -- artificial neural network or $k$-means/kNN algorithms. This model can be applied on a test set of \NCP to reproduce corrections schemes for a system with a fixed parameter $\gamma$.  
The generated pulses \DCP can be integrated with different values of parameter $\gamma$. The fidelity of the resulting operator with the target operator provides the efficiency of the approximation. In the other words, we test how efficient is \DCP generated for some $\gamma$ when we apply it to a system with different drift strength.

\begin{figure}[ht!]
  \centering{
    \includegraphics[width=0.5\textwidth]{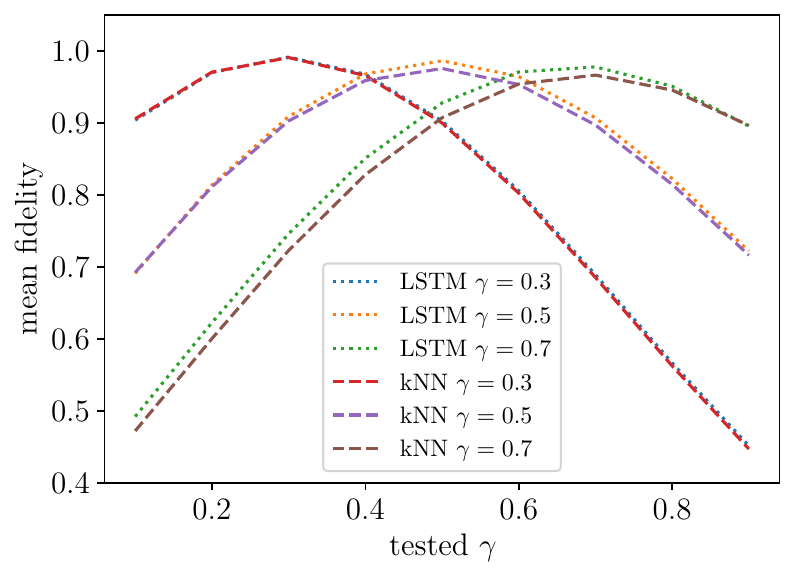}
    \caption{Reference point of generalization for LSTM and kNN.}
    \label{fig:reference_points_gam03_05_07}}
\end{figure}

The reference points for kNN and LSTM models are presented in Fig.~\ref{fig:reference_points_gam03_05_07}. As expected from the results of Task I, LSTM has higher efficiency than kNN near the value of $\gamma$ used to construct the reference point. 

\subsubsection{Generalization efficiency}

In the numerical experiments testing the generalization efficiency the input is a vector of triples, each  of the form $(h_x(t), h_z(t),\gamma)$. We denote inputs as (\NCP, $\bar{\gamma}$), where $\bar{\gamma}$ denotes the vector with all elements equal to $\gamma$ and the appropriate dimension. The constant vector with all elements equal to $\gamma$ can be interpreted as the additional third dimension in the time series. Hence, in the case of artificial neural networks it is possible to utilize similar architecture as in the case of Task~I. The difference is that input time point is a triple $x_t=(\NCP_x(t),\NCP_z(t),\gamma$) (see Appendix~\ref{sec:LSTM}).

During the training phase we utilize the values of the drift strength
from some set $A$. Next, during the test phase, we are given other drift strength values $\gamma\notin A$. In the numerical experiments presented below we use two sets of parameters for training, $A_1=\{0.1, 0.3, 0.5\}$ and $A_2=\{0.5, 0.7, 0.9\}$. During the testing phase  we use parameters  from set $B=A_1\cup A_2\cup\{0.2,0.4,0.6,0.8\}$.

In the case of clustering, the training and the testing phases of Task II are executed as follows.

\begin{enumerate}
  \item \textbf{Generating training set} Generate Haar random unitary matrices. Based on them, generate \NCP and \DCP for $\gamma\in\{0.1,0.3,0.5,0.7,0.9\}$. For each $\gamma$ value we use 3000 training samples, what gives us 9000 training samples for set $A_1$ and 9000 for set $A_2$.

  \item \textbf{Training} Create approximation of the correction scheme as in Fig.~\ref{fig:execution_dcp_various_gamma}
  
  \begin{enumerate}
    \item For LSTM network, we first create pairs (\NCP, $\bar{\gamma}$), with values of $\gamma\in A_1 \cup A_2$. Next we train two LSTM networks independently, the first on the training data corresponding to set $A_1$ and the second on the data corresponding to set $A_2$. Each input (\NCP, $\bar{\gamma}$) has a corresponding target \DCP. Cost function is the same as in Task I.

    \item For the geometrical method, first we perform the clustering algorithm with the number of clusters equal to 500, on $\CCP = \DCP -\NCP$ obtained from \DCP for $\gamma \in A_1$ and $A_2$.
    
    One should note that we apply the kNN on the flattened \NCP, corresponding to \CCP from the clustering. However, we include additional information about the drift by adding an element equal to $\gamma$ multiplied by some large number.
    This changes the meaning of the additional dimension and separates the vectors on the subspace spanned by the added element. In our numerical experiments, this number is equal to 1000. In kNN we choose $k=4$.
     
  \end{enumerate}

  \item \textbf{Testing} Generate 2000 Haar random unitary matrices and generate corresponding vectors of control pulses,~\NCP.\\ 
  Testing phase is based on the joining \NCP with testing value of gamma $\gamma\in B$ in a suitable manner for each algorithm. Next, using ML algorithm we generate \nnDCP, and integrate it as in Eq.~\eqref{eq:integrate_superoperator}. After that we compute the fidelity error between the obtained operator and the target operator.

\end{enumerate}


One should note that each pair (\NCP, $\bar{\gamma}$) corresponds to a different \DCP. Hence, due to the different number of drift values used in both phases, we have different number of samples in the training and testing phases.

\subsubsection{Comparison of the results}

\begin{figure}[ht!]
  \centering    
  
  \subfigure[Comparison of the efficiency of generalization obtained using 
  kNN with reference points for $\gamma = 0.3$ and 0.5.]{
    \includegraphics[width=0.45\textwidth]{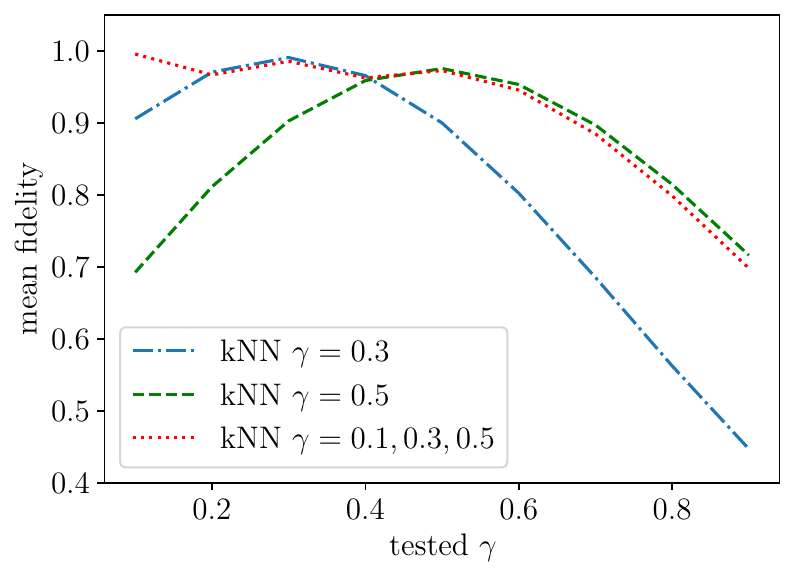}
    \label{fig:generalization_kMeanskNN_gam01_03_05}}
  \subfigure[Comparison of the efficiency of generalization obtained using 
  kNN with reference points for $\gamma = 0.5$ and 0.7.]{
    \includegraphics[width=0.45\textwidth]{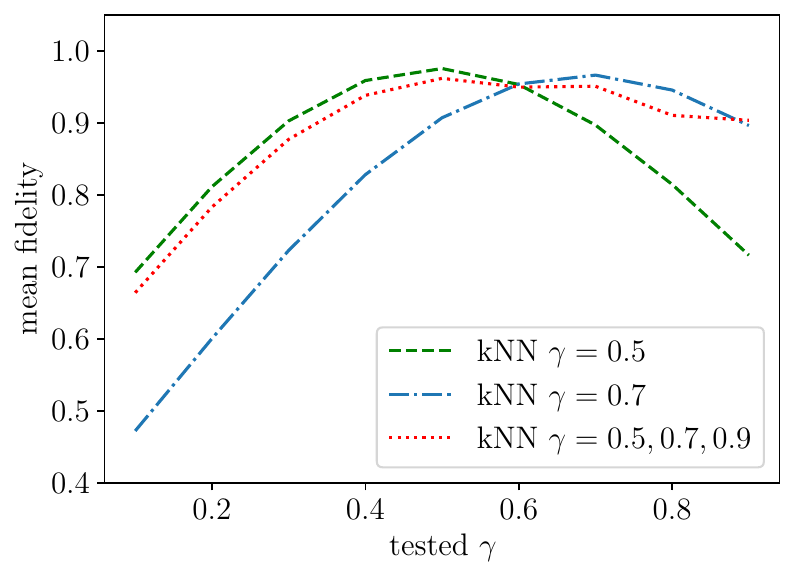}
    \label{fig:generalization_kMeanskNN_gam05_07_09}}

  \subfigure[Comparison of the efficiency of generalization obtained using 
LSTM with reference points for $\gamma = 0.3$ and 0.5.]{
  \includegraphics[width=0.45\textwidth]{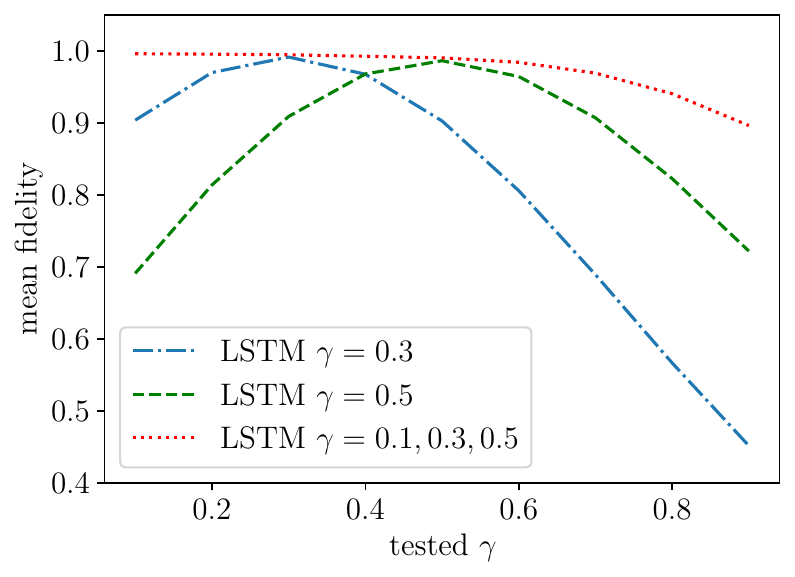}
  \label{fig:generalization_LSTM_gam01_03_05}}
\subfigure[Comparison of the efficiency of generalization obtained using 
LSTM with reference points for $\gamma = 0.5$ and 0.7.]{
  \includegraphics[width=0.45\textwidth]{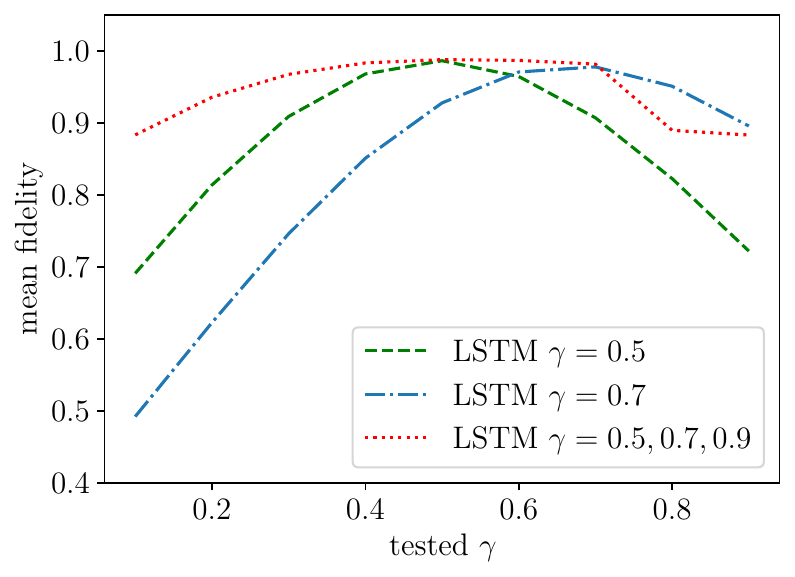}
  \label{fig:generalization_LSTM_gam05_07_09}}

\caption{
  Generalization abilities of kNN and LSTM methods. Figs.~\subref{fig:generalization_kMeanskNN_gam01_03_05} and 
  \subref{fig:generalization_kMeanskNN_gam05_07_09} represent efficiency of control pulses obtained using kNN. Figs.~\subref{fig:generalization_LSTM_gam01_03_05} and \subref{fig:generalization_LSTM_gam05_07_09} represent efficiency of control pulses obtained using LSTM. Each plot was made for different values of drift strength used during the training phase.
  In all cases, the training set has size 3000, and the testing set has size 2000.}
\label{fig:training_with_gamma}
\end{figure}

The above procedures enable us to analyse the changes in the efficiency for both methods in the function of the disturbance strength. As one can see in Fig.~\ref{fig:generalization_kMeanskNN_gam01_03_05}, the $k$--means/kNN has three local maxima, which correspond to the values of drift strength on which the algorithm was trained. Therefore, this algorithm has noticeable drops in the interpolation. One can conclude that the clustering trained on the values from $A$ is as good as the clustering trained on a single value.

Moreover, for kNN trained with $\gamma \in A_1$, the extrapolation is worse than the reference point for $\gamma=0.5$. For kNN trained with $\gamma \in A_2$,
the ability to generalize for other $\gamma$ is also limited. In this case, this might be caused by the presence of outliers in the training data (see Fig~\ref{fig:generalization_kMeanskNN_gam05_07_09}).
This suggests that this method does not utilize the information about 
$\gamma$ parameter. The additional $\gamma$ in the input is not utilized and 
the algorithm obtains similar results as the algorithm without $\gamma$ in the 
input. 

The situation is different in the case of utilizing of LSTM network, which displays the ability to generalize the correction scheme using information about parameter $\gamma$. This effect can be observed in Fig.~\ref{fig:generalization_LSTM_gam01_03_05}. As one can see, LSTM has high efficiency in the neighbourhood of the training points.
 
Moreover, for LSTM trained with $\gamma \in A_1$, the extrapolation is better than the reference point for $\gamma=0.5$ and especially for $\gamma=0.3$. Thus, one can conclude that LSTM has the ability to generalize for other $\gamma$.

One should note that the ability of generalization does not depend on the  values of $\gamma$. The efficiency of extrapolation depends only on the difference between the testing values and the training values.  LSTM method has better generalization abilities when trained with small values of $\gamma$ and used to generate control pulses for large values, as well as in the case where we use larger values for training and extrapolate to smaller values.
%

\section{Concluding remarks}\label{sec:fianl}

The results presented in this paper demonstrate that both neural networks and 
geometrical methods provide good correction schemes and 
enable counteracting the undesired interaction present in the physical system. 
The performed numerical experiments suggest that there are some types of nontrivial quantum 
control tasks for which simple geometrical data representation and processing 
is as good as much more sophisticated machine learning models. We have also observed that the 
same tasks can become much more demanding when interpolating and extrapolating 
with respect to the system parameters is expected.

One should note that both methods studied in this work have their specific 
advantages and
disadvantages. Numerical results suggest that recurrent neural networks have 
the ability to generalize their predictions with respect to the system  
parameters.
This can be seen in the presented numerical experiments where the network trained on a limited set of gamma values also has good results for gamma values which were absent in the training process. Moreover, artificial neural networks are useful in the sense of approximation function as the trained network is a deterministic map from \NCP to \DCP. Because of this, one can examine a variation, continuity, and other mathematical features of this correction scheme~\cite{ostaszewski18approximation}.

On the other hand, the application of clustering shows that this repair scheme 
on its own, when generalization is not considered, can be compressed to a 
relatively small number of corrections. This demonstrates that the continuous 
process of quantum control can be represented be a relatively small number of 
representative control pulses. Such method has the efficiency of 
approximation similar as in the case of the recurrent neural networks. 
Results presented in this paper suggest that the machine learning approach 
based on simple, geometrical methods can provide a viable approach for the 
analysing of the locally affine transformations on the space of control pulses.

The analysis presented in this paper can be seen as a step in the process of compiling quantum operations into control pulses. Machine learning methods could be utilized to improve the quality of the obtained control pulses. This would require the application of the presented methods for quantum control methods used in current quantum computers.

\section*{Acknowledgements}
  MO acknowledges
  support from Polish National Science Center scholarship 2018/28/T/ST6/00429. JAM
  acknowledges support from Polish National Science Center grant
  2014/15/B/ST6/05204. Authors would like to thank Daniel Burgarth and Leonardo Banchi for discussions about quantum control, Adam Glos, Bartosz Grabowski, and Wojciech Masarczyk for discussions concerning the details of LSTM architecture, and Izabela Miszczak for reviewing the manuscript. Numerical calculations
  were  possible thanks  to the support  of PL-Grid  Infrastructure.

%


%
\appendix

\section{LSTM as an approximation of the correction scheme}
\label{sec:LSTM}

As an approximation of the correction scheme which takes into account time 
series structure of data, we have used bi-directional long
short-term memory (biLSTM) units~\cite{hochreiter1997long}. LSTM is a special 
type of artificial neural network, which processes its hidden states in 
recurrent (cyclic) way.

\begin{figure}[ht!]
  \begin{tikzpicture}[>=latex,
  ->,shorten >=1pt,
  cell/.style={
    align=center,
    rectangle, 
    rounded corners=2mm, 
    draw
  },
  input/.style={
  },
  hidden/.style={
  },
  subcell/.style={
    align=center,
    rectangle, 
    draw,
    minimum width = 1.cm,
    minimum height = 0.7cm,
    rounded corners=1mm
  },
  arrow/.style={
    rounded corners=.2cm
  }
  ]
  
  \pgfmathsetmacro{\yg}{-1.4}
  \pgfmathsetmacro{\yh}{-2.4}
  \pgfmathsetmacro{\yi}{-3.0}
  \pgfmathsetmacro{\as}{0.15}
  
  \node [cell, minimum height =1.cm, minimum width=5cm] (cell) at 
  (2.3,0){
      
        $
      c_t = c_{t-1}\circ \textrm{f}_t + \tilde{c}_t\circ \textrm{i}_t,
      $
      
  } ;
  \node [subcell] (forget) at (-1.2,\yg) {$
    \textrm{f}_t =\mySigma(x_tV^\textrm{f}+s_{t-1}W^\textrm{f})
    $};
  \node [subcell] (input) at (2.3,\yg) {  $
    \textrm{i}_t = \mySigma(x_tV^\textrm{i}+s_{t-1}W^\textrm{i})
    $};
  \node [subcell] (candid) at (6.1cm,\yg) {$
    \tilde{c}_t = \tanh(x_tV^\textrm{o}+s_{t-1}W^\textrm{o})
    $};
  \node [subcell] (output) at (10cm,\yg) {$
    \textrm{o}_t = \mySigma(x_tV^\textrm{o}+s_{t-1}W^\textrm{o})
    $};
  
  \node [cell, minimum height =1.cm, minimum width=1cm] (hidden) at (10,0) 
  {
      

      $
      s_t=\tanh (c_t)\circ \textrm{o}_t,
      $
  };
  
  \node [hidden] (prevh0) at  (-2,\yh) {$s_{t-1}$};
  \node [hidden] (prevhf) at  (0,\yh) {};
  \node [hidden] (prevhi) at  (2,\yh) {};
  \node [hidden] (prevhc) at  (4,\yh) {};
  \node [hidden] (prevho) at  (6,\yh) {};
  
  \node [input] (prevct) at (-2,0) {$c_{t-1}$};
  
  \node [input] (prev0) at  (-2,\yi) {$x_{t}$};
  \node [input] (prevf) at  (0,\yi) {};
  \node [input] (previ) at  (2,\yi) {};
  \node [input] (prevc) at  (4,\yi) {};
  \node [input] (prevo) at  (6,\yi) {};
  
  \draw [arrow] (prevct) -- (cell);
  
  \draw [arrow] (prevh0 -| prevhf) ++ (-1.4,0) -| ($(forget.south)-(\as,0)$);
  \draw [arrow] (prevh0 -| prevhf) ++ (-1.4,0) -| ($(input.south)-(\as,0)$);
  \draw [arrow] (prevh0 -| prevhf) ++ (-1.4,0) -| ($(candid.south)-(\as,0)$);
  \draw [arrow] (prevh0 -| prevhf) ++ (-1.4,0) -| ($(output.south)-(\as,0)$);
  
  \draw [arrow] (prev0 -| prevf) ++ (-1.4,0) -| ($(forget.south)+(\as,0)$);
  \draw [arrow] (prev0 -| prevf) ++ (-1.4,0) -| ($(input.south)+(\as,0)$);
  \draw [arrow] (prev0 -| prevf) ++ (-1.4,0) -| ($(candid.south)+(\as,0)$);
  \draw [arrow] (prev0 -| prevf) ++ (-1.4,0) -| ($(output.south)+(\as,0)$);
  
  \draw [arrow] (forget) -- ($(cell.south)-(2,0)$);;
  \draw [arrow] (input) -- (cell);
  \draw [arrow] (candid) -- ($(cell.south)+(2,0)$);
  \draw [arrow] (output) -- (hidden);
  
  \node [hidden] (outc) at (7,1.1) {$c_t$};
  \node [hidden] (outh) at (10,1.1) {$s_t$};
  
  \draw [arrow] ($(cell.east)-(0,\as)$) -- ($(hidden.west)-(0,\as)$);
  \draw [arrow] ($(cell.east)+(0,\as)$ -| hidden) ++ (cell) -| (outc);
  
  \draw [arrow] (hidden) -- (outh);
  
  \draw [arrow] (outc) -- (7,1.5) [rounded corners] -- (2.4,1.5);
  \draw [arrow] (outh) -- (10,1.9) [rounded corners] -- (2.4,1.9);;
  
  \node (dots1) at (2.0,1.5) {$\dots$};
  \node (dots2) at (2.0,1.9) {$\dots$};
  
  \draw [arrow] (1.5,1.5) -- (-3.0,1.5) [rounded corners] -- (-3.0,1) -- 
  (-3.0,0) -- (prevct) [rounded 
  corners];
  \draw [arrow] (hidden) -- (12,0);
  \draw [arrow] (1.5,1.9) -- (-3.4,1.9) [rounded corners] -- (-3.4,\yh)  -- 
  (prevh0) [rounded corners];

  \end{tikzpicture}
  \caption{Structure of LSTM network used in the numerical  experiments (from~\cite{ostaszewski18approximation}). }
\end{figure}

All gates have common input 
$x_t$, 
the current input, $s_{t-1}$, the previous hidden state, and the weight 
matrices $W$ and $V$. In $c_t$ (cell state) and $s_t$ (hidden state), operation $\circ$ is 
element-wise multiplication of two vectors. 
Because of sigmoid function, $\mySigma$, outputs $i_t$, $f_t$ and $o_t$ of 
gates are 
vectors with elements from $[0,1]$ interval. This construction enables the 
maintenance the information, which is propagated to next time step. 
The bi-directional version of LSTM analyses the input sequence in both directions -- 
forwards and 
backwards~\cite{schuster1997bidirectional,graves2005bidirectional}. As the 
input, it takes  $[h(t_1), h(t_2),\ldots ,h(t_n)]$ and 
$[h(t_{n}), h(t_{n-1}),\ldots ,h(t_1)]$.

The utilized architecture consists of three layers of bidirectional LSTM with 
the number of hidden units respectively 200, 250, and 300. The output of the last 
LSTM unit is merged as a element-wise sum of forward LSTM output and backward 
LSTM product. As the final step, the fully connected perceptron layer maps a 300-dimensional vector to the origin size. As the optimization method we used Adam 
optimizer~\cite{kingma2014adam} with learning rate $0.0001$. As the cost 
function we chose mean squared error between the prediction of the network and 
the target. Numerical experiments were 
performed using the \texttt{TensorFlow} 
library~\cite{abadi2016tensorflow,tensorflow}.

\section{Function approximation by clustering method}\label{sec:generation-kmeans}

\subsection{Clustering correction algorithm}
\label{sec:clusterring_corr}

\begin{algorithm}[H]
  \caption{\texttt{Clustering correction algorithm}}
  \begin{algorithmic}[1]
    \Require {A set of affine corrections $C = \{c_i = b_i - a_i\}$, number of clusters $k$}
    \State{(Optional) Approximation of \CCP vectors
      
      {We choose the type of approximation (one from Eqs.~\eqref{eq:sine},\eqref{eq:poly3}, or \eqref{eq:poly4}) and change the space of considered data \ie we express every $\textrm{c}_i$ by a vector of suitable approximation coefficients -- $\textrm{coeffs}_i$}}

    \State{Clustering of the set of $C$  
      
      {Apply $k$-means algorithm on the set of corrections $C = \{{c_i}\}_i$ or on $\textrm{coeffs}_i$. As every $\textrm{coeffs}_i$ corresponds to $c_i$, the result is the set of $k$ clusters 
        $C_1,C_2,\ldots,C_k$ and corresponding them labels $y_i$}}
    \State{Calculate output corrections as the mean correction 
      \begin{equation}\bar{C_i}\leftarrow\frac{\sum_{{c}\in C_i}{c}}{|C_i|}\end{equation}}
    
%
    \Return{ (labels -- $\{y_i\}$, corrections -- $\{\bar{C_i}\}$}
    
  \end{algorithmic}
  \label{alg:knn_approx}
\end{algorithm}

Output of the above algorithm and training set of arguments gives us the approximation of the map of \NCP to \DCP. To get correction of the new point \NCP we need to develop a method for deciding which
correction $\{\bar{C}_j\}$ should be 
used. For this purpose we utilize kNN algorithm, which will be 
fitted on the set of training  $\{\NCP_i\}_i$ with labels obtained during the clustering of $\{\CCP_i\}_i$.
Prediction of kNN on test \NCP gives us the most probable cluster $j$, and in
result the correction $\bar{C_j}$ which should be used to 
generate suitable  \nnDCP $\approx$ \DCP. 

It should be stressed that input data in our numerical experiments are not row vectors \ie \NCP, \DCP and \CCP in above algorithm are matrices. Thus, for $k$-means and kNN algorithm, they are flattened to raw vectors.

\subsection{Approximation schemes}
\label{sec:approx_fncs}
In this approach, we also analyse if the data can be compressed via approximation by elementary functions
\begin{enumerate}[a)]
  \item sinusoid
  \begin{equation}
  f_{sinusoid}(x)=\alpha_1\sin(\alpha_2 x +  \alpha_3) + \alpha_4\sin(\alpha_5 x + \alpha_6) + \alpha_7,
  \label{eq:sine}
  \end{equation}
  with $\alpha_1 >\alpha_4$,
  \item polynomial of the third degree (poly3)
  \begin{equation}
  f_{poly3}(x)=\alpha_1x^3 + \alpha_2x^2 + \alpha_3x + \alpha_4,
  \label{eq:poly3}
  \end{equation}
  \item polynomial of the forth degree (poly4)
  \begin{equation}
  f_{poly4}(x)=\alpha_1x^4 + \alpha_2x^3 + \alpha_3x^2 + \alpha_4x + \alpha_5.
  \label{eq:poly4}
  \end{equation}
\end{enumerate}

Here all $\alpha_i$ are real parameters used in the 
approximation. In the numerical experiments, \CCP will be the signal which we will try to approximate, and thus $x=\CCP$. Clustering is made on the parameters $\alpha_i$ corresponding to particular \CCP.

\subsection{k nearest neighbours algorithm}
\label{sec:kNN}
Let us assume that we have a given training set $(x_1,y_1),\ldots, (x_n,y_n)$, where 
$x_i\in \mathbb{R}^m$ are data points, and $y_i\in \{1,\ldots ,l\}$ are  
corresponding labels. The core idea of the algorithm is as follows. For each new sample 
$x_{n+1}$ we take $k$ nearest samples from the training set, and based on the 
majority of votes, we assign label $y_{n+1}$. 
\begin{algorithm}[H]
  \caption{$k$NN}
  \begin{algorithmic}[1]
    \Require {Hyperparameter $k$, training data points $(x_1,y_1),\ldots, (x_n,y_n)$, and test sample $x$.}
    \For{ $i=1,\dots,n$ }
    \State{$d_i\leftarrow ||x_i-x||$}
    \EndFor
    \State{Compute multi-set of labels $S$ corresponding to $k$ smallest distances $d_i$}
    \For{ $j=1,\dots,l$ }
    \State{$e_j\leftarrow\frac{|S\cap \{j\}|}{k}$}
    \EndFor
    \State{}
    \Return $\arg\max_je_j$
  \end{algorithmic}
  \label{alg:kNN}
\end{algorithm}
One should note that it is possible to construct many variants of this algorithm by changing the applied  norm. In the the scope of this work, we use Euclidean norm. Moreover, it is worth to note that this method provides an example of \emph{instance-based method}, \ie it does not create any generalization rules, but compares each sample to the training set.

\end{document}